\documentclass[prc,showpacs]{revtex4}

\usepackage{amssymb}
\usepackage{bm}
\usepackage{graphicx}

\begin{document}

\title{Weak interaction rates for Kr and Sr waiting-point nuclei under rp-process
conditions}

\author{P. Sarriguren}
\email{sarriguren@iem.cfmac.csic.es}

\affiliation{Instituto de Estructura de la Materia, CSIC, Serrano
123, E-28006 Madrid, Spain}

\begin{abstract} 

Weak interaction rates are studied in neutron deficient Kr and Sr 
waiting-point isotopes in ranges of densities and temperatures relevant 
for the $rp$ process. The nuclear structure is described within a 
microscopic model (deformed QRPA) that reproduces not only the 
half-lives but also the Gamow-Teller strength distributions recently 
measured. The various sensitivities of the decay rates to both density 
and temperature are discussed. Continuum electron capture is shown to 
contribute significantly to the weak rates at $rp$-process conditions.

\end{abstract}


\maketitle

\section{Introduction}
\label{intro}

Nuclear physics is an essential piece in the present understanding
of many astrophysical processes related to the energy generation and
nucleosynthesis in stellar scenarios. The properties of both stable 
and exotic nuclei play different roles at different stages of stellar 
evolution. While the former are relevant to describe steady regimes,
the latter are implicated in the description of explosive events.
Network calculations and astrophysical models rely on the quality of
the input information, in particular, on the quality of the nuclear 
input. Unfortunately, the experimental information available for 
exotic nuclei is still very short and thus, most of the astrophysical 
simulations of violent phenomena must be based on nuclear model 
predictions of limited quality and accuracy. This is the case of the 
X-ray bursts, generated by a thermonuclear runaway in the 
hydrogen-rich environment of an accreting compact object (typically 
a neutron star) which is fed from a binary companion (typically a 
red giant). The ignition takes place on the surface of the neutron 
star at high densities ($\rho$) and temperatures ($T$), and 
eventually peak conditions of $T=$1-3 GK and 
$\rho=10^6$-$10^7$ g cm$^{-3}$ are reached \cite{schatz,wormer}. 
The mechanism leading to X-ray bursts is the rapid proton ($rp$) 
capture process \cite{schatz,wormer,wallace}, which is characterized 
by proton capture reaction rates that are orders of magnitude faster 
than any other competing process, in particular $\beta$-decay. It 
produces rapid nucleosynthesis on the proton-rich side of stability 
toward heavier proton-rich nuclei and the energy observed in X-ray 
bursts. 

Nuclear reaction network calculations (the set of differential equations 
for the various isotopic abundances) are performed 
\cite{schatz,wormer,wallace,woosley} to follow the time 
evolution of the isotopic abundances, to determine the amount of energy
released by nuclear reactions, and to find the reaction path for the 
$rp$ process. In general, the reaction path follows a series of fast 
proton-capture reactions until further proton capture is inhibited. 
Then the reaction flow has to wait for a relatively slow $\beta$-decay 
and the respective nucleus is called a waiting point (WP). The half-lives 
of the WP nuclei along the reaction path determine the time scale of the 
nucleosynthesis process and the produced isotopic abundances.
In this respect, the weak decay rates of neutron-deficient 
medium-mass nuclei under stellar conditions play a relevant role to 
understand the $rp$ process. The nuclear structure problem involved in 
the calculation of these rates must be treated in a reliable way. 
In particular, this implies that the nuclear models should be able to 
describe at least the experimental information available on the decay 
properties (Gamow-Teller strength distributions and $\beta$-decay 
half-lives) measured under terrestrial conditions. Although these decay 
properties may be different at the high $\rho$ and $T$ existing in 
$rp$-process scenarios, success in their description under terrestrial 
conditions is a requirement for a reliable calculation of the weak 
decay rates in more general conditions.
With this aim in mind, we study in this letter the dependence of the
decay rates on both  $\rho$ and $T$, using a nuclear model that has 
been tested successfully and reproduces the experimental information 
available on both bulk and decay properties of medium-mass
nuclei very reasonably. 
This model is the quasiparticle random phase approximation (QRPA). 
Here, we focus our attention to the deformed WP isotopes $^{72,74}$Kr 
and $^{76,78}$Sr, where the Gamow-Teller (GT) strength distributions 
have been measured with high accuracy by $\beta$-decay over most of 
the $Q$-window \cite{piqueras,poirier,nacher} and are well reproduced 
by theoretical calculations based on deformed QRPA \cite{sarri_last}.
While the half-lives give only a limited information of the decay
(different strength distributions may lead to the same half-life), 
the strength distribution contains all the information. 
It is the first time that the weak decay rates under stellar $rp$ 
conditions can be studied using a nuclear structure model that 
reproduces the GT strength distributions and half-lives under 
terrestrial conditions \cite{sarri_last,sarri_wp}.

There are several distinctions between terrestrial and stellar decay 
rates caused by the effect of high  $\rho$ and $T$.
The main effect of $T$ is related to the thermal population of
excited states in the decaying nucleus, accompanied by the corresponding 
depopulation of the ground states. It has been shown \cite{moller} that 
the weak decay rates of excited states can be significantly different
from those of the ground state, enhancing the total decay rates. A case 
by case consideration is needed.
Another effect related to the high  $\rho$ and $T$ comes from the fact 
that atoms in these scenarios will be completely ionized and the 
electrons will be no longer bound to the nuclei, but forming a degenerate
plasma obeying a Fermi-Dirac distribution. This opens the possibility 
for continuum electron captures ($cEC$).
These effects make weak interaction rates in the stellar interior 
sensitive functions of $T$ and $\rho$, with $T=1.5$ GK and 
$\rho=10^6$ g cm$^{-3}$, as the most relevant conditions for the $rp$ 
process \cite{schatz}.

\section{Weak decay rates in stellar scenarios}
\label{weak}

The general formalism to calculate weak interaction rates in stellar 
environments as a function of $\rho$ and $T$ was introduced in the 
pioneering work of Fuller, Fowler, and Newman \cite{ffn}.
Further improvements have come mainly on the nuclear structure 
aspect, either from the Shell Model \cite{langanke,pruet} or from 
QRPA \cite{nabi}.

\begin{figure}[h]
\centering
\includegraphics[width=120mm]{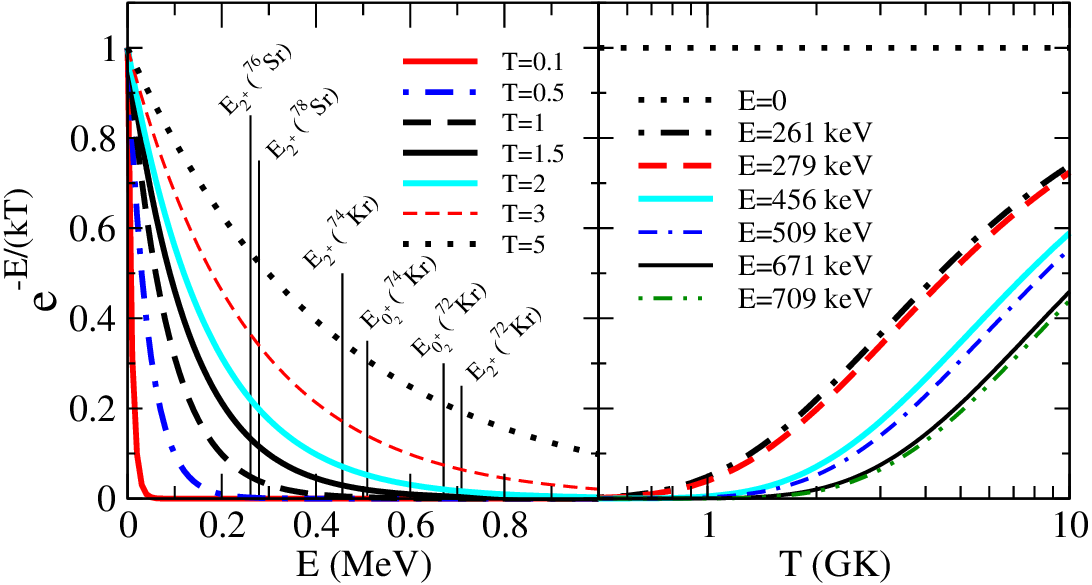}
\caption{Population coefficients as a function of the excitation energy 
(left) and temperature (right). The energies of the low-lying excited states 
of the isotopes considered in this work are shown explicitly.}
\label{fig_ekt}
\end{figure}

The decay rate of the parent nucleus is given by \cite{ffn}

\begin{equation}
\lambda = \sum_i \lambda_i\, \frac{2J_i+1}{G} e^{-E_i/(kT)} \, ; \qquad
G=\sum_i (2J_i+1) e^{-E_i/(kT)} \, ,
\label{population}
\end{equation}
where $J_i\, (E_i)$ is the angular momentum (excitation energy)
of the parent nucleus state $i$, and thermal equilibrium is assumed. 
In principle, the sum extends over all populated states in the parent 
nucleus up to the proton separation energy. However, since the range 
of $T$ for the $rp$-process peaks at $T=1.5$ GK ($kT\sim 300$ keV),
only a few low-lying excited states are expected to contribute 
significantly in the decay. Specifically, we consider in this work 
all the (collective) low-lying excited states below 
1 MeV \cite{bouchez,ensdf} (exception made of the 
$E_{4^+}=782$ keV in $^{78}$Sr),
$E_{0^+}=671$ keV and $E_{2^+}=709$ keV in $^{72}$Kr;
$E_{0^+}=509$ keV and $E_{2^+}=456$ keV in $^{74}$Kr;
$E_{2^+}=261$ keV in $^{76}$Sr; and $E_{2^+}=279$ keV in $^{78}$Sr.
Then, all these states are considered to be thermally populated
although for $T=1.5$ GK the population is 
negligible beyond 500 keV. 
In Fig.~\ref{fig_ekt} one can see the population coefficients 
in Eq. (\ref{population}) as a function of the excitation energy (left) 
and temperature (right). As an example, the population of the lowest
of these states ($E_{2^+}=261$ keV in $^{76}$Sr) at $T$=1.5 GK is 
about $12\%$.

The decay rate for the parent state $i$ is given by

\begin{equation}
\lambda _i = \sum_f \lambda_{if}\, ,
\end{equation}
where the sum extends over all the states in the final nucleus reached
in the decaying process. The rate $\lambda_{if}$ from the initial state
$i$ to the final state $f$ is given by

\begin{equation}
\lambda _{if} = \frac{\ln 2}{D}  B_{if}\Phi_{if} (\rho,T)\, ,
\end{equation}
where $D=6146 \,{\rm s}$. This expression is decomposed into
a nuclear structure part $B_{if}$ and a phase space factor $\Phi_{if}$.

$B_{if}$ contains the transition probabilities
for allowed Fermi and GT transitions,

\begin{equation}
B_{if}=B_{if}(GT)+ B_{if}(F)\, ,
\end{equation}
with

\begin{equation}
B_{if}(GT)= \frac{1}{2J_i+1} \left( \frac{g_A}{g_V} \right)_{\rm eff} ^2 
\langle f || \sum_k \sigma^k t_{\pm}^k || i \rangle ^2 \, ,
\end{equation}
$(g_A/g_V)_{\rm eff} = 0.74 (g_A/g_V)_{\rm bare}$ 
is an effective quenched value.

\begin{equation}
B_{if}(F)= \frac{1}{2J_i+1}\langle f || \sum_k t_{\pm}^k || i \rangle ^2
=t(t+1)-t_{zi}t_{zf} \, ,
\end{equation}
where $t$ is the isospin, which is assumed to be a good quantum number.
For $\beta^+$ transitions and very neutron deficient nuclei with $Z>N$, 
the superallowed Fermi transition to the isobaric analog state has 
to be taken into account. However, for $N>Z$ nuclei, Fermi transitions
are not possible except for isospin impurities. 
The discussion is therefore limited to the decay via allowed GT transitions.

The theoretical formalism used here to calculate the GT strengths is based
on the QRPA. The quasiparticle
basis corresponds to a deformed selfconsistent Skyrme Hartree-Fock 
calculation with SLy4 \cite{sly4} force and pairing correlations treated 
in BCS approximation. The residual interactions are spin-isospin forces 
in both particle-hole and particle-particle channels. The initial and
final states in the laboratory frame are expressed in terms of
the intrinsic states using the Bohr-Mottelson factorization \cite{bm},
which is a very good approximation for well deformed nuclei. 
The effect of all these ingredients (deformation, pairing, residual forces) 
on the decay rates has been studied elsewhere \cite{sarr1} and their 
relevance in this mass region has been stressed.
More details of the formalism can be found in Ref. \cite{sarr2}. 
The present formalism has been shown to provide a good description of the 
decay properties of nuclei in the mass region $A\sim 70$.
In particular, the parameters for residual forces used here are the same
used in Ref. \cite{sarri_last}, where good agreement was obtained with 
the experimental GT strength distributions and $\beta$-decay half-lives.
The success of this theoretical formalism in reproducing terrestrial decay 
rates supports its application to the calculation of the weak interaction 
rates in stellar matter.

The $\beta^+$ and $cEC$ phase space integrals are given by

\begin{equation}
\Phi_{if}=\Phi^{cEC}_{if}+\Phi^{\beta^+}_{if}\, ,
\end{equation}
where
\begin{eqnarray}
\Phi^{cEC}_{if}&=&\int_{\omega_\ell}^{\infty} \omega
\sqrt{\omega^2-1} (Q_{if}+\omega)^2
F(Z,\omega) \nonumber \\
&& \times S_e(\omega) \left[ 1-S_{\nu}(Q_{if}+\omega)\right] d\omega \, ,
\label{phiec}
\end{eqnarray}
for continuum electron capture, and
\begin{eqnarray}
\Phi^{\beta^+}_{if}&=&\int _{1}^{Q_{if}} \omega \sqrt{\omega^2-1}  
(Q_{if}-\omega)^2 F(-Z+1,\omega) \nonumber \\
&& \times \left[ 1-S_p(\omega)\right]
\left[ 1-S_{\nu}(Q_{if}-\omega)\right] d\omega \, ,
\label{phib}
\end{eqnarray}
for positron emission. In these expressions
$\omega$ is the total energy of the electron (positron) in units of $m_ec^2$ 
and $F(Z,\omega)$ is the Fermi function \cite{gove}
that takes into account the distortion of the $\beta$-particle wave function 
due to the Coulomb interaction.

\begin{figure}[h]
\centering
\includegraphics[width=120mm]{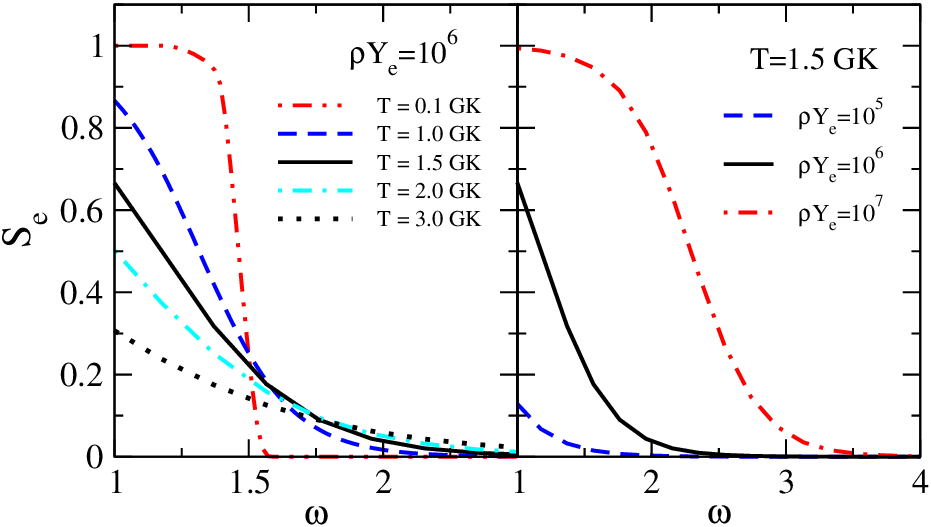}
\caption{Electron distributions as a function of the total
electron energy for various $T$ at fixed  $\rho$ (left panel) and for various  
$\rho$ at fixed $T$ (right panel).}
\label{fig_se}
\end{figure}

\begin{equation}
Q_{if}=\frac{1}{m_ec^2}\left( M_p-M_d+E_i-E_f \right)
\end{equation}
is the total energy available in the decay in units of $m_ec^2$. It is 
written in terms of the nuclear masses of parent ($M_p$) and daughter 
($M_d$) nuclei and their excitation energies $E_i$ and $E_f$, respectively.
In the $cEC$ factor, the lower integration limit is given by
${\omega_\ell}=1$ if $Q_{if}> -1$, or ${\omega_\ell}=|Q_{if}|$ if 
$Q_{if}< -1$. $S_e$, $S_p$, and $S_\nu$, are the electron, positron, and 
neutrino distribution functions, respectively. Its presence inhibits or 
enhances the phase space available. In $rp$ scenarios the commonly accepted 
assumptions \cite{schatz,ffn,langanke,pruet,nabi} state that $S_\nu=0$, since 
neutrinos can escape freely from the interior of the star without blocking 
their emission in the capture or decay processes. Positron distributions 
become only important at higher $T$ ($kT > 1$ MeV) when they appear via 
pair creation. At the temperatures considered here we take $S_p=0$.
The electron distribution is described by a Fermi-Dirac distribution

\begin{figure}[h]
\centering
\includegraphics[width=120mm]{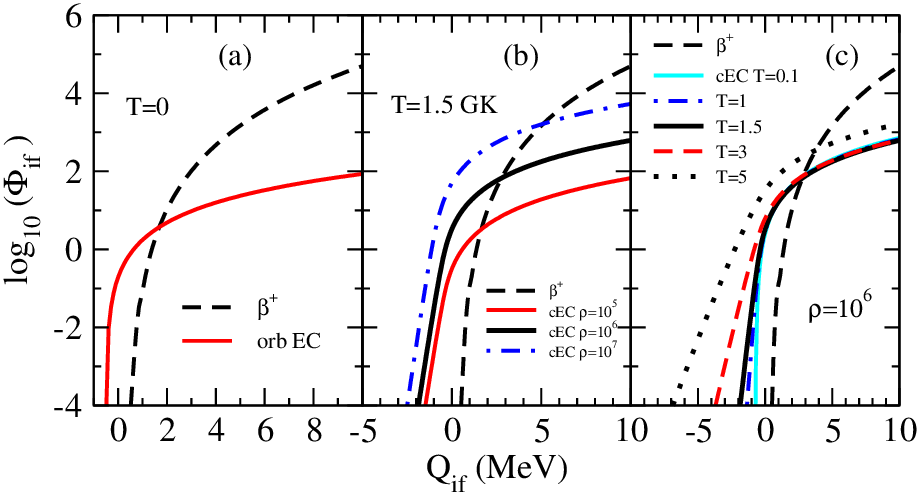}
\caption{Phase space factors for Kr as a function of the total energy 
available in the decay for various $T$ (GK) and $\rho$ (g cm$^{-3}$). 
Electron capture factors are calculated
from orbital electrons in the case $T=0$ (a) and from continuum electrons in 
the case of nonzero $T$ (b,c).}
\label{fig_phi}
\end{figure}

\begin{equation}
S_{e}=\frac{1}{\exp \left[ \left(\omega -\mu_e\right)/(kT)\right] +1} \, ,
\end{equation}
assuming that nuclei at these temperatures are fully ionized and 
the electrons are not bound to nuclei.
The chemical potentials $\mu_e$ are determined from $\rho$ and $T$. 
Tables for them can be found in Ref. \cite{ffn} for selected choices 
of $T$ and $\rho$.

Fig.~\ref{fig_se} shows $S_e(\omega)$ for various temperatures and 
electron densities $\rho Y_e$ ($\rho$ is the baryon density and $Y_e$ 
stands for the ratio of the electron to the baryon numbers) around 
the peak $rp$ conditions. One can see in Fig.~\ref{fig_se} that $S_e$ 
becomes more important at higher densities, as well as the relevance 
of electrons with high energies as $T$ increases. These features 
establish the relative importance of the $cEC$ rates as $\rho$ and 
$T$ change.

Fig.~\ref{fig_phi} contains the phase space factors for $\beta^+$ decay
and electron capture as a function of the total energy available $Q_{if}$.
In (a) one can see the usual phase factors at zero $T$, that is the  
$\beta^+$ in Eq.~(\ref{phib}), which is independent of $T$ and $\rho$ and 
the orbital electron capture calculated as described in Ref. \cite{gove}.
In (b) and (c) one can also see the $cEC$ factors as a function of $\rho$ 
and $T$. In general, they increase with $Q_{if}$ and thus the decay rates
are more sensitive to the strength $B_{if}$ located at low excitation 
energies of the daughter nucleus ($E_f$). It is also interesting to notice 
the relative importance of both  $\beta^+$ decay and electron capture. 
The latter is always dominant at sufficiently low $Q_{if}$, or 
correspondingly at sufficiently high excitation energies $E_f$.
In this work the effects of bound-state electron capture are not included 
since they are insignificant over most of the ranges of $T$ and 
$\rho$ considered.

\begin{figure}[h]
\centering
\includegraphics[width=130mm]{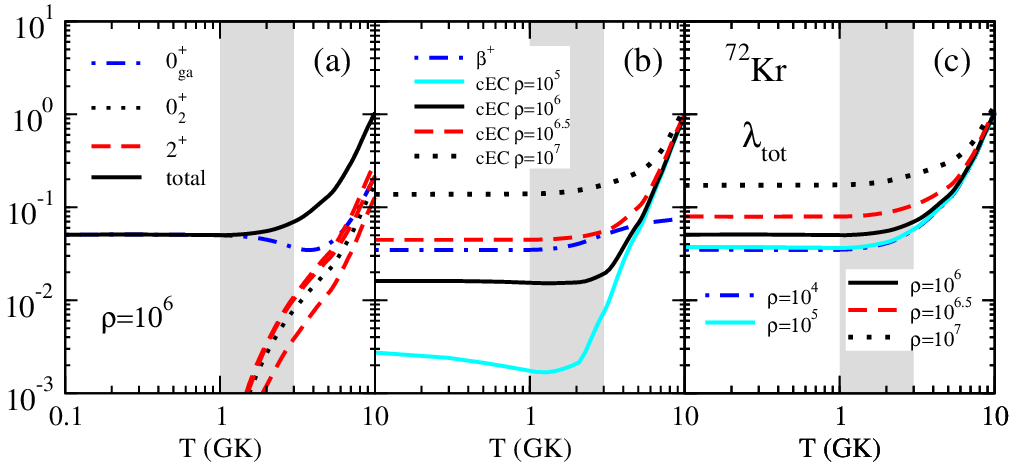}
\caption{Weak decay rates $\lambda$ (s$^{-1}$) for $^{72}$Kr 
as a function of $T$ for various $\rho$.}
\label{fig_rates_kr}
\end{figure}

\begin{figure}[h]
\centering
\includegraphics[width=130mm]{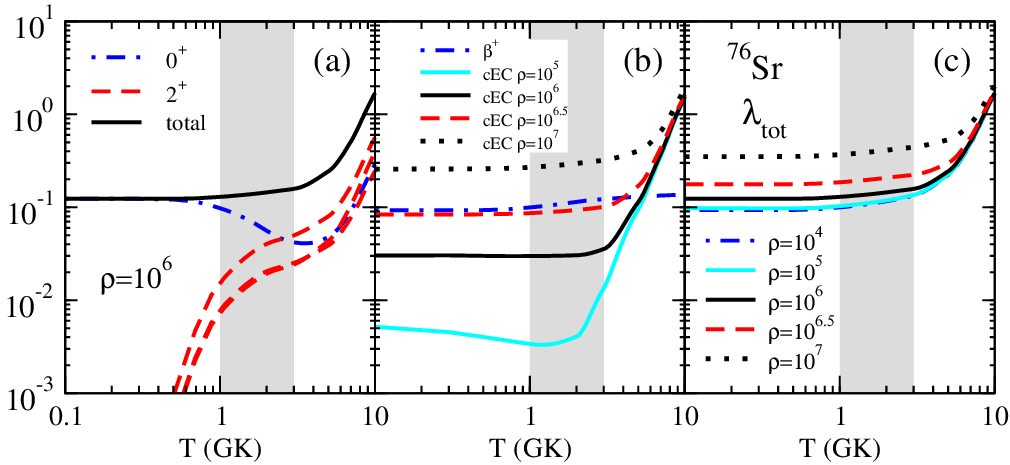}
\caption{Same as in Fig. \ref{fig_rates_kr}, but for $^{76}$Sr.}
\label{fig_rates_sr}
\end{figure}

\section{Results}
\label{results}

Fig.~\ref{fig_rates_kr} shows the 
decay rates versus $T$ on the example of $^{72}$Kr. In panel (a) one can 
see the decomposition of the total rate into their contributions from the 
decay of the the ground state $0^+_{\rm gs}\rightarrow 1^+$ and from the 
decay of the excited states $0^+_2\rightarrow 1^+$ and 
$2^+ \rightarrow 1^+,2^+,3^+$, which are negligible except at very 
high $T$. Panel (b) contains the decomposition of the rates into their  
$\beta^+$ and  $cEC$  components evaluated at different densities. 
One can see that for densities larger than $\log_{10}(\rho)\sim 6.5$, $cEC$ 
becomes dominant at any $T$. For lower densities,  $\beta^+$ rates are 
larger than $cEC$ at low $T$, but lower at sufficiently high $T$.
Since the  $\beta^+$ rate is practically independent of both $\rho$ and 
$T$, the total rate in panel (c) is determined by  $\beta^+$ at low $T$ 
and $\rho$, and by $cEC$ otherwise. The gray area is the relevant range 
$T$=1-3 GK for the $rp$ process. Fig.~\ref{fig_rates_sr} shows the same 
rates as in Fig. \ref{fig_rates_kr}, but for $^{76}$Sr.
The only difference worth to mention with respect to $^{72}$Kr is the
relatively more important contributions of the $2^+$ excited state, as
seen in panel (a), which is the result of a larger population of this 
state due to its lower excitation energy.

\begin{figure}[h]
\centering
\includegraphics[width=120mm]{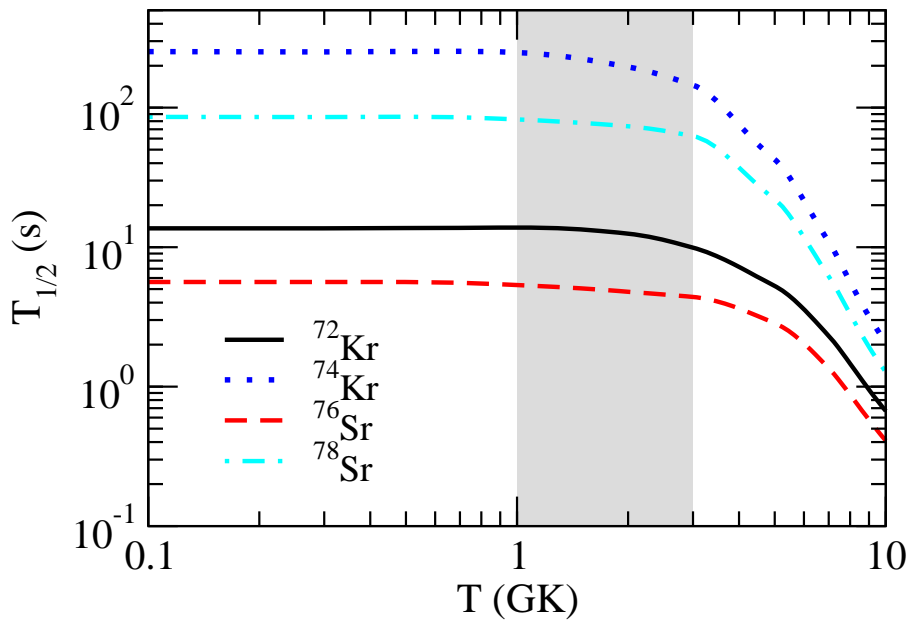}
\caption{Half-lives as a function of $T$ at a
density $\rho =10^6$ g cm${^{-3}}$ for Kr and Sr isotopes.}
\label{fig_t12}
\end{figure}

Finally, Fig.~\ref{fig_t12} contains the half-lives ($T_{1/2}=\ln 2/\lambda$),
including $\beta^+$ and $cEC$ contributions, as a function of $T$ at a 
fixed $\rho = 10^6$ for the various isotopes. One can see the decrease 
of $T_{1/2}$ as $T$ increases. The decrease starts to be significant 
beyond $T=3$ GK, which is outside of the relevant temperature range for 
the $rp$ process. One should notice that the orbital electron capture 
has been ignored in this work in favor of continuum electron capture, 
but both should be considered to obtain a smooth transition toward 
terrestrial conditions. 
The calculation at $T=0$ with  orbital electron capture contributions 
can be found in Ref. \cite{sarri_last}, where comparison
with experiment is also shown.

\section{Conclusions}
\label{conclusions}

In conclusion, the weak decay rates of Kr and Sr WP nuclei have been 
investigated at temperatures and densities where the $rp$ process
takes place. It has been the first time that stellar weak rates have 
been calculated in this mass region with a nuclear structure model 
that reproduces both the half-lives and the more demanding GT strength 
distributions measured with high accuracy from $\beta$-decay experiments.
In this letter we have 
analyzed the relevant ingredients to describe the rates in a reliable 
way. We have studied: 1) The contributions to the decay rate coming 
from excited states in the parent nucleus which are populated as $T$ 
raises. All the relevant states below 1 MeV have been included.
It is found that below $T=3$ GK their effect can be neglected, 
and thus, the decay from the ground state is already a good 
approximation for $rp$ processes. This conclusion is in agreement 
with previous studies \cite{schatz,langanke,petrovici}. Nevertheless,
one should pay special attention to the cases where the $2^+$ excited 
states are particularly low in energy because their contributions 
can be competitive at these temperatures;
2) The effect of the continuum electron capture rates. It is
found that $cEC$ rates are enhanced as $T$ and $\rho$ increase and 
they become comparable to the $\beta^+$ decay rates at $rp$ peak 
conditions. At slightly larger values of $T$ and $\rho$, $cEC$ 
dominates over $\beta^+$ decay. This point is important since 
$cEC$ contributions have been neglected \cite{schatz} in earlier 
evaluations of weak decay rates at $rp$ conditions.

\noindent {\bf Acknowledgments}

This work was supported by Ministerio de Ciencia e Innovaci\'on
(Spain) under Contract No.~FIS2008--01301.

\end{document}